\documentclass{article}
\pdfoutput=1 
\usepackage{graphicx}
\usepackage[english]{babel}
\usepackage{hyperref}
\usepackage{upgreek} 
\usepackage{wrapfig}   
\usepackage{amsmath} 
\usepackage{cite} 
\newcommand{\ALICE}{\mbox{ALICE}} 
\newcommand{\PbPb}{\mbox{Pb-Pb}}
\newcommand{\Vzero}{\ensuremath{\mathrm{V}^0}}
\newcommand{\Lam}{\ensuremath{\Lambda}}
\newcommand{\Sig}{\ensuremath{\Sigma}}
\newcommand{\Xii}{\ensuremath{\Xi}} 
\newcommand{\Ome}{\ensuremath{\Omega}}

\newcommand{\rsnn}{\ensuremath{\sqrt{s_{\mathrm{NN}}}}} 
\newcommand{\mT}{\ensuremath{m_{\mathrm{T}}}} 
\newcommand{\pT}{\ensuremath{p_{\mathrm{T}}}} 
\newcommand{\dedx}{\ensuremath{\mathrm{d}E/\mathrm{d}x}}
\newcommand{\mev}{\mbox{MeV}}
\newcommand{\mevc}{\mbox{MeV$/c$}}
\newcommand{\gevc}{\mbox{GeV$/c$}}
\newcommand{\gevcc}{\mbox{GeV$/c^2$}}
\newcommand{\ie}{i.\,e.}   
\newcommand{\eg}{e.\,g.}
\newcommand{\vs}{vs.}
\newcommand{\etal}{\textit{et~al.}} 
\newcommand{\mytitle}{Femtoscopic
  \texorpdfstring{p\Lam}{Proton-Lambda}
  Correlations in \PbPb\ Collisions at
  \texorpdfstring{$\rsnn$}{sqrt(sNN)}
  = 2.76~TeV with ALICE}

\makeatletter
\@ifpackageloaded{hyperref}{%
  \hypersetup{%
    pdftitle = {\mytitle{}},
    pdfsubject = {Proceedings for invited talk at WPCF 2014},
    pdfkeywords = {proton lambda correlations femtoscopy ALICE
      Pb-Pb 2.76~TeV},
    pdfauthor = {\textcopyright\ Hans Beck}
  }
}{}
\makeatother

\title{
  \includegraphics[width=0.35\textwidth]{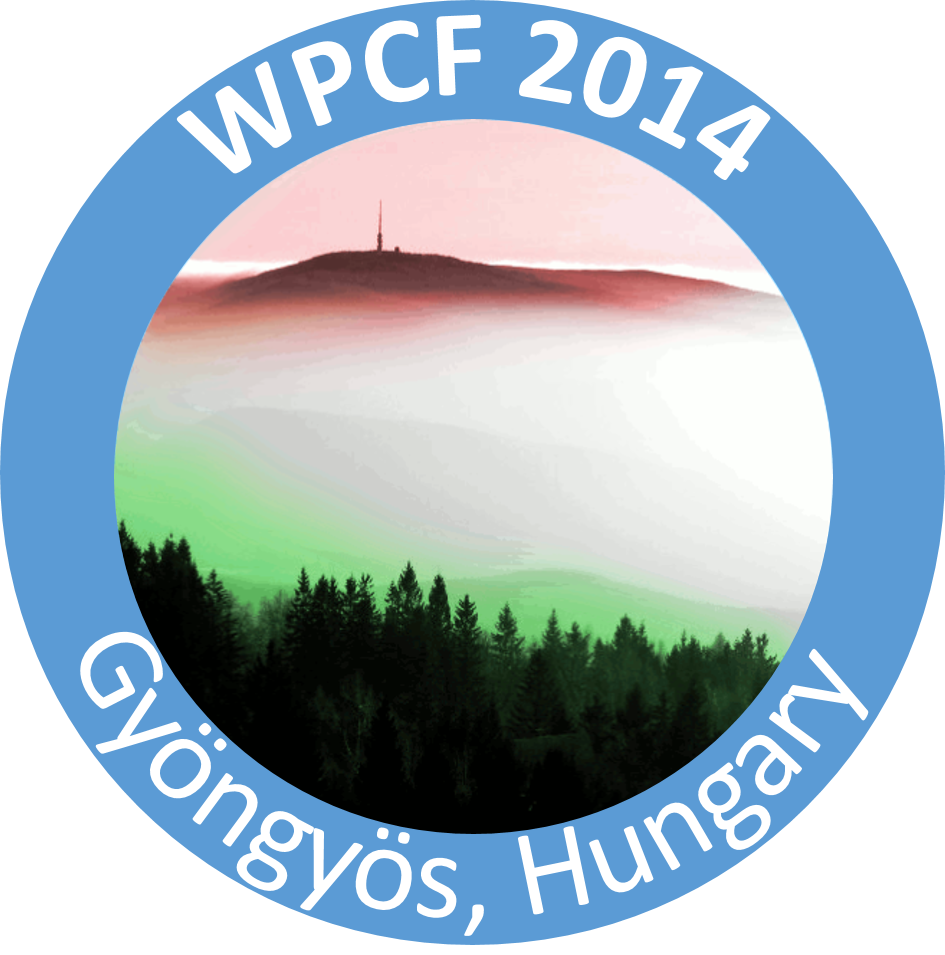}\\[1cm]
  \mytitle
}
\author{
  Hans Beck\footnote{email: Hans.Beck@cern.ch}\\[1ex]
  Institut f\"{u}r Kernphysik, Goethe-Universit\"{a}t, Frankfurt,
  Germany\\[1ex]
  for the \ALICE\ Collaboration
}

\begin{document}

\maketitle

\begin{abstract}
Two-particle correlations at small relative momenta give insight into
the size of the emitting source. Of particular interest is the test of
hydrodynamic models which predict a universal, apparent decrease of
the extent of the system with increasing transverse mass \mT\ as a
consequence of the strong radial flow in heavy-ion collisions at LHC
energies. This contribution presents a study of correlations for
protons and \Lam\ particles in \PbPb\ collisions at $\rsnn =
2.76$~TeV measured with \ALICE. The investigated particle species expand
the experimental reach in \mT\ due to their high rest mass. Residual
impurities in the samples from misidentification and contributions
from feed-down are corrected using data-driven techniques. 
Correlation
functions are obtained in several centrality classes and \mT\
intervals at large \mT\ and show the expected decrease in volume
of the strongly interacting fireball 
for more peripheral collisions. We observe the decrease of the source size with \mT,
predicted by hydrodynamical calculations,
out to $\langle \mT \rangle = 2.18~\gevcc$.
\end{abstract}
\section{Introduction}
Femtoscopy aims at measuring the size of the particle emitting
source. Experimentally, this is achieved by studying two-particle
correlations at small relative momenta. If the size of the system is
large, the emissions of the two particles will likely be
substantially separated in space.
Therefore, no interaction or
symmetrization will take 
place and the momenta will be
uncorrelated. If -- on the other hand -- 
the extent of the source is small, the vicinity of the
emanation points will trigger the species-specific interplay with a
characteristic dependence of the correlation on the momentum
difference. Given that the two-particle interaction is known, a source
size can be inferred from the correlation function.

No
Coulomb or quantum-statistical effects take place in the p\Lam\ system
and the strong 
interaction parameters are sufficiently well known from, \eg, 
bubble chamber experiments. 
A typical reaction for studying the
p\Lam\ final-state interaction is the production of \Lam\
baryons by shooting a beam of negatively charged kaons on a hydrogen
target as done with the Saclay bubble chamber~\cite{saclay}. The \Lam\ 
hyperon -- created via $\mathrm{K}^- + \mathrm{p} \rightarrow \Lam +
\uppi^0$ -- can subsequently scatter elastically off another
proton. Measuring the cross-section of the elastic process
differentially \vs\ the excess energy directly quantifies the
attractive final-state strong interaction, which turns out to be
comparable to the nucleon-nucleon one.

Based on the model by Lednick\'y and
Lyuboshits~\cite{LedLyuSovJNuclPhys}, the sensitivity of the p\Lam\ 
correlation function to the volume 
of the hot and dense medium created in heavy-ion collisions was first
explored in~\cite{wangprattprl}. The correlation was found to be
affected in height and shape by the size of the source. Furthermore,
the interdependence of the two momenta was shown to keep its 
susceptibility to a change in source size for radii larger than 4~fm
-- an advantage over the Coulomb-depleted proton-proton
correlations.

A particularly interesting subject is the investigation of the reaction
dynamics of the strongly interacting matter generated in \PbPb\
collisions. The large pressure gradients give rise to an expansion of
the medium. The resulting collective velocity competes with the
thermal velocity $\sqrt{T/\mT}$; consequently the \mT\ dependence of
the radii probes the dynamics of the source. At low \mT, no strong
collective motion is present and the particles will be correlated over
the full extent of the source. At high \mT\ in contrast, the radial
flow introduces a positive correlation between the emission point and
the particle's momentum. Looking only at pairs of particles with a small
momentum difference, the couple's constituents will originate
from the same region, which is smaller than the geometrical size of
the fireball, leading to the apparent shrinking of the particle source
with \mT. To probe the high \mT\ regime, it is experimentally
beneficial to investigate particles with a high rest mass. So far, the p\Lam\
system is the heaviest studied system~\cite{Starplam,Starpp,NA49plam}.

\section{Particle Selection}
The analyzed \PbPb\ data was taken with
\ALICE~\cite{ALICE-JINST}. The T0 and V0 scintillator arrays
provided an event-trigger signal and an estimate for the collision
centrality. The main detectors for particle tracking in ALICE are the
Inner Tracking System, Time Projection Chamber (TPC), and Transition
Radiation Detector. Particle identification for this analysis was
performed by the TPC and the Time Of Flight (TOF) detector. The
vanishing $\mu_{B}$ at the LHC entails that matter and anti-matter are
produced in equal abundance.
While the correlation functions for pairs of p\Lam\ and
$\overline{\mathrm{p}}\overline{\Lam}$ were obtained separately, the
selections for particles discussed in the following hold true for
their charge conjugates as well.

The proton selection depended on the reconstructed momentum and used
the TPC and/or TOF. 
The resolution of the truncated specific energy loss measurement
(\dedx) can be parametrized with a
Gaussian of width $\sigma$. 
The TPC provided a separation of the most
probable \dedx\ measurement for protons to the one of any other
species of more than four $\sigma$ up to $p = 0.75~\gevc$.
Above this kinematic restriction, the expected \dedx\ for protons and
the ultra-relativistic electrons becomes too similar, which impeded an
unambiguous particle identification by the TPC alone. 
For higher
momenta up to $p = 5.0~\gevc$, TOF allowed for an unambiguous proton
identification~\cite{ALICE-Performance}. With the chosen criteria a
proton purity -- defined as the number of protons over the
number of all selected particles including misidentified tracks -- 
above 99\% was achieved over the full dynamic span.

\begin{figure}[htb]
  \vspace{-0.2cm}
  \begin{center}$
    \begin{array}{cc}
      \includegraphics[width=0.488\textwidth]{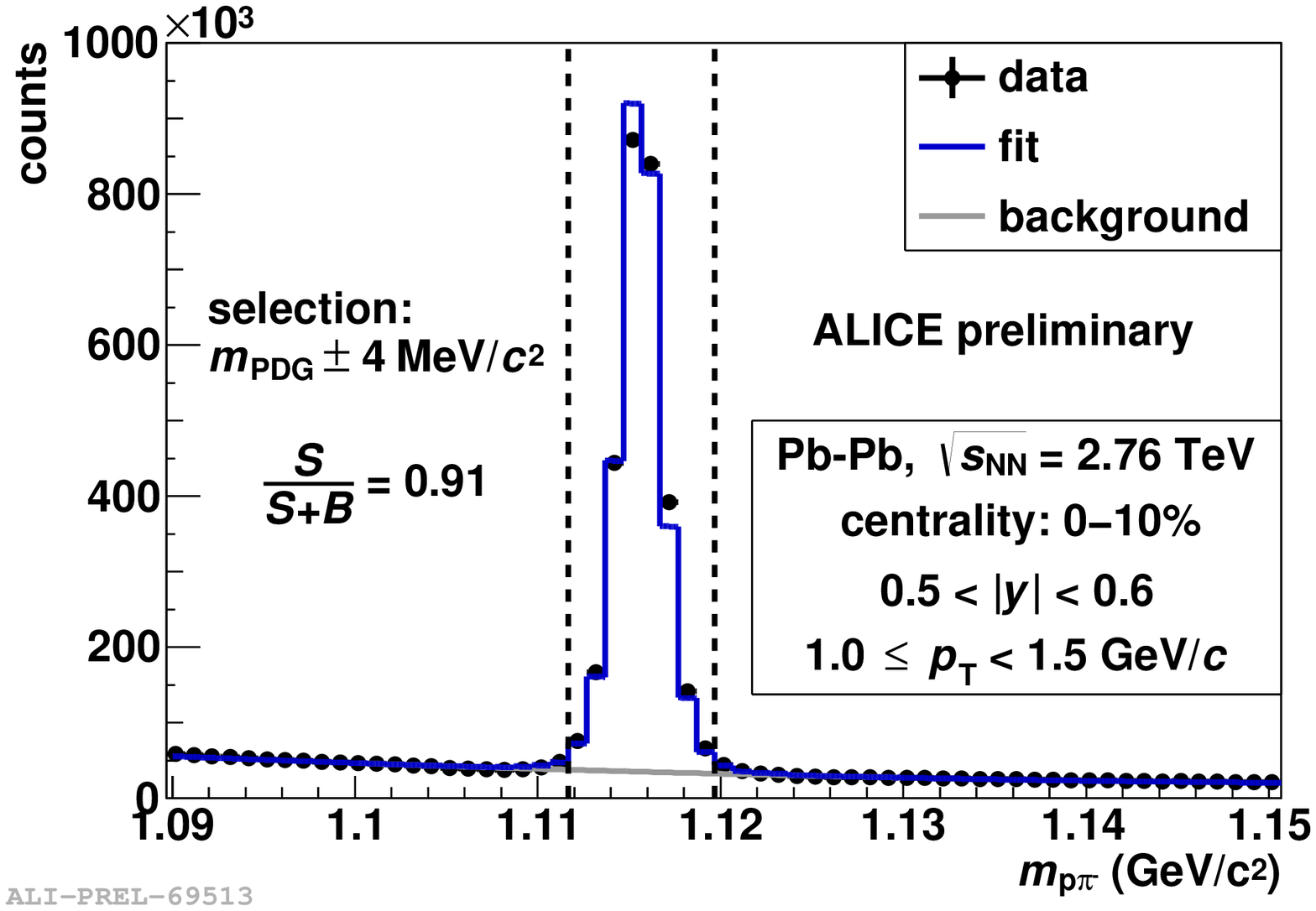} &
      \includegraphics[width=0.488\textwidth]{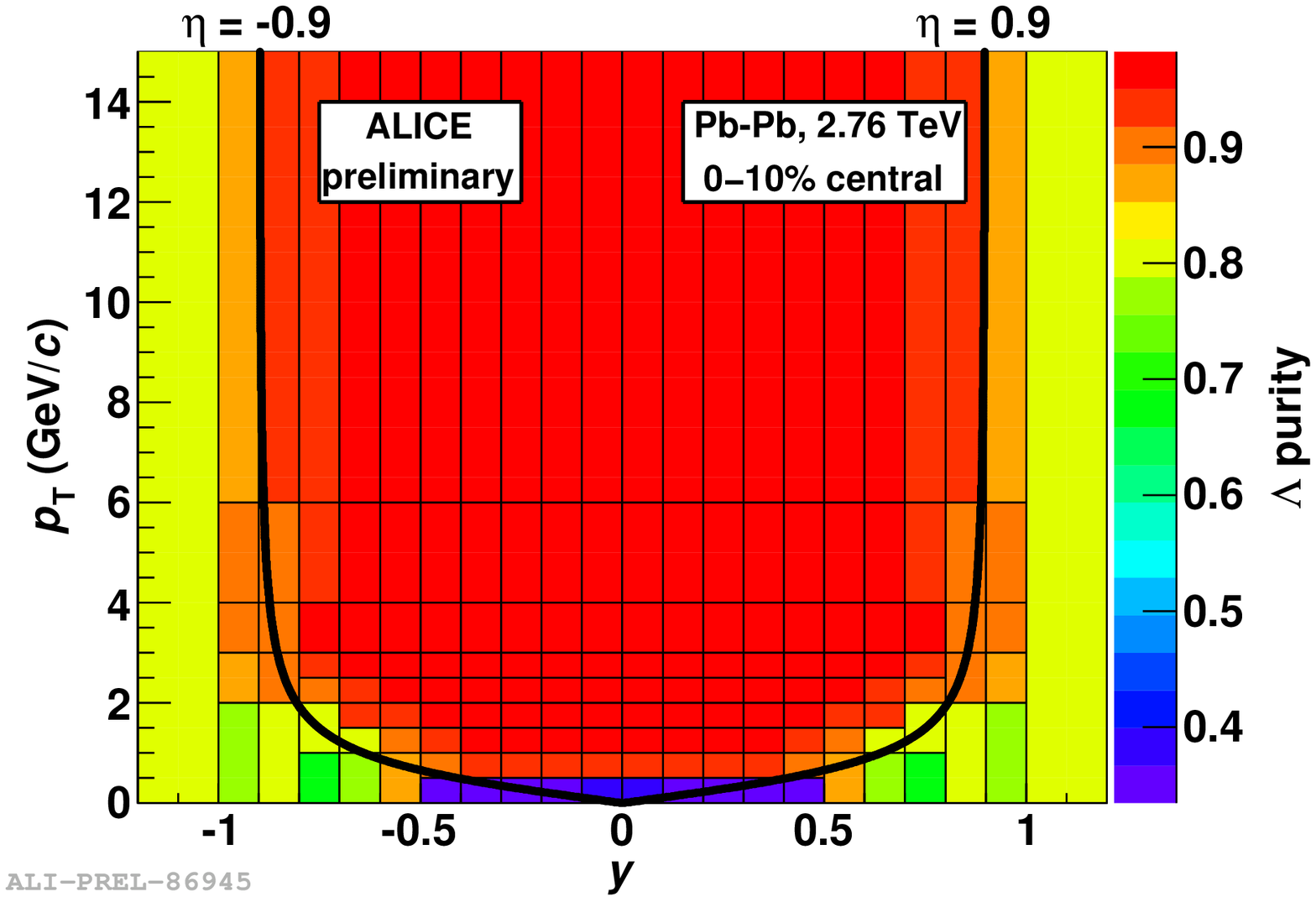}
    \end{array}$
  \end{center}
  \vspace{-0.5cm}
  \caption{Left: Invariant mass spectrum of \Lam\ candidates in
    an exemplary phase-space bin with a fit using a Monte-Carlo
    template for the signal shape.
  Right: \Lam\ purity as a function of rapidity and transverse
  momentum for the 10\% most central events.}
  \label{fig:lampur}
\end{figure}

The \Lam\ selection was based upon a \Vzero\ topology finder, which
reconstructs the charged decay $\Lam \rightarrow \mathrm{p} +
\uppi^-$; the hyperon is identified via the invariant mass. 
Fig.~\ref{fig:lampur} (left) shows a fit to the invariant mass
distribution of all \Lam\ candidates employing a Monte-Carlo
template for the signal shape in an exemplary phase-space bin with a
high yield, namely $0.5 < |y|< 0.6$ and $1.0 \le \pT~(\gevc) <
1.5$, for the 10\% most central events. The parametrization allows to
determine the purity of the sample -- defined as the ratio of picked
\Lam\ particles to all taken \Vzero\ vertices -- which amounts in this case to 
$\mathrm{pur}_\Lam = 91\%$ within the window in invariant mass of
$\pm 4~\mev$ around the PDG value. Determining the purity enables one to
correct the correlation function for the uncorrelated background in a
following step. Fig.~\ref{fig:lampur} (right) shows the evolution of
the \Lam\ purity as a function of rapidity and transverse 
momentum for the 10\% most central events.
Protons knocked-out from an interaction with the detector material resemble
displaced tracks from weak particle decay vertices. 
A contribution of protons from material manifests itself in a
degradation of the purity for $\pT < 0.5~\gevc$; the effect is absent
for the case of $\overline{\Lam}$. For higher transverse momenta,
the purity is 81\% for $|y| > 1.0$, within $|\eta| < 0.9$ it is better
than 90\%. 

\section{Feed-Down Determination}
\begin{figure}[htb]
  \begin{center}$
    \begin{array}{cc}
      \hspace{-0.02\textwidth}
      \includegraphics[width=0.53\textwidth]{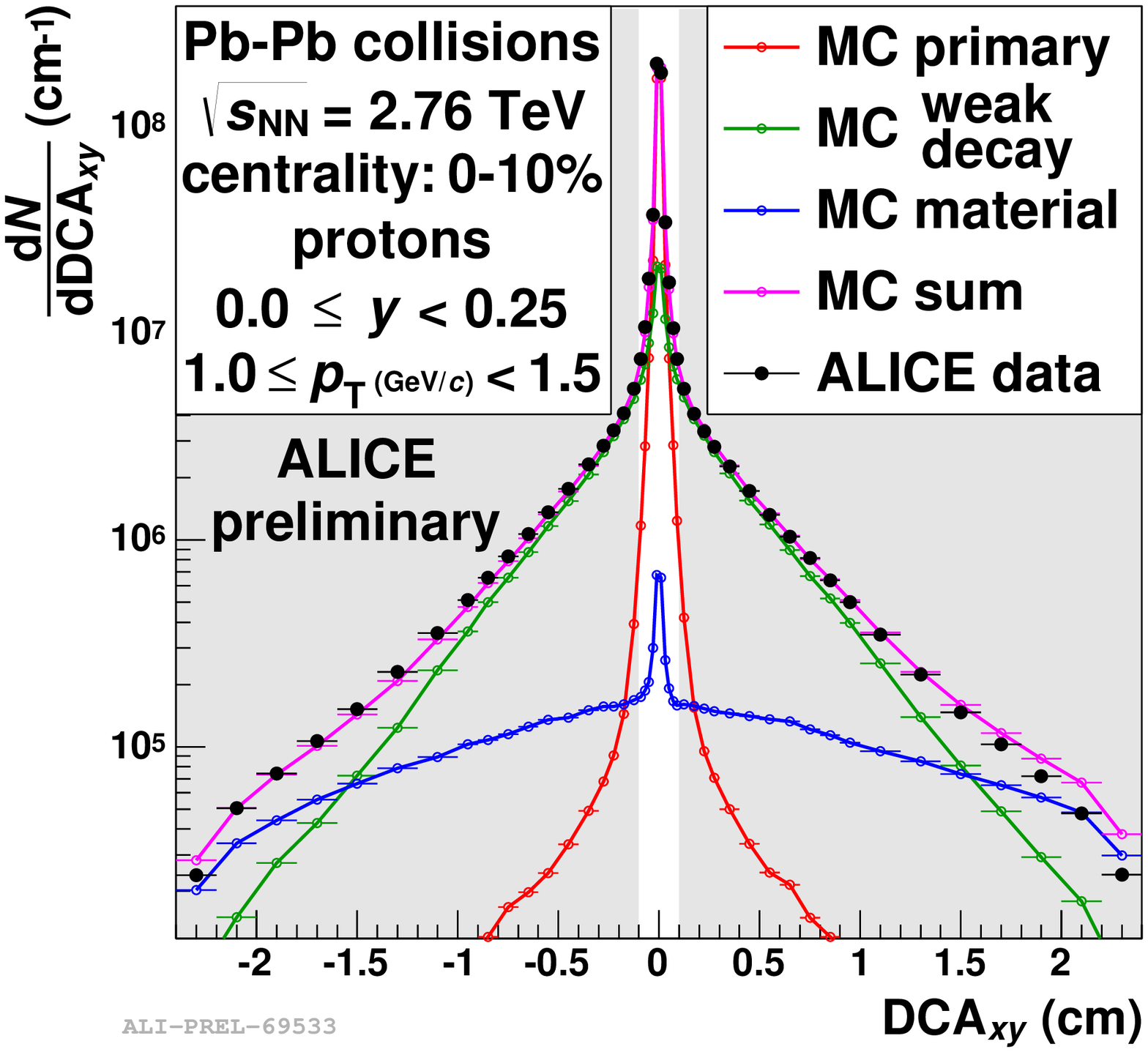} &
      \hspace{-0.03\textwidth}
      \includegraphics[width=0.488\textwidth]{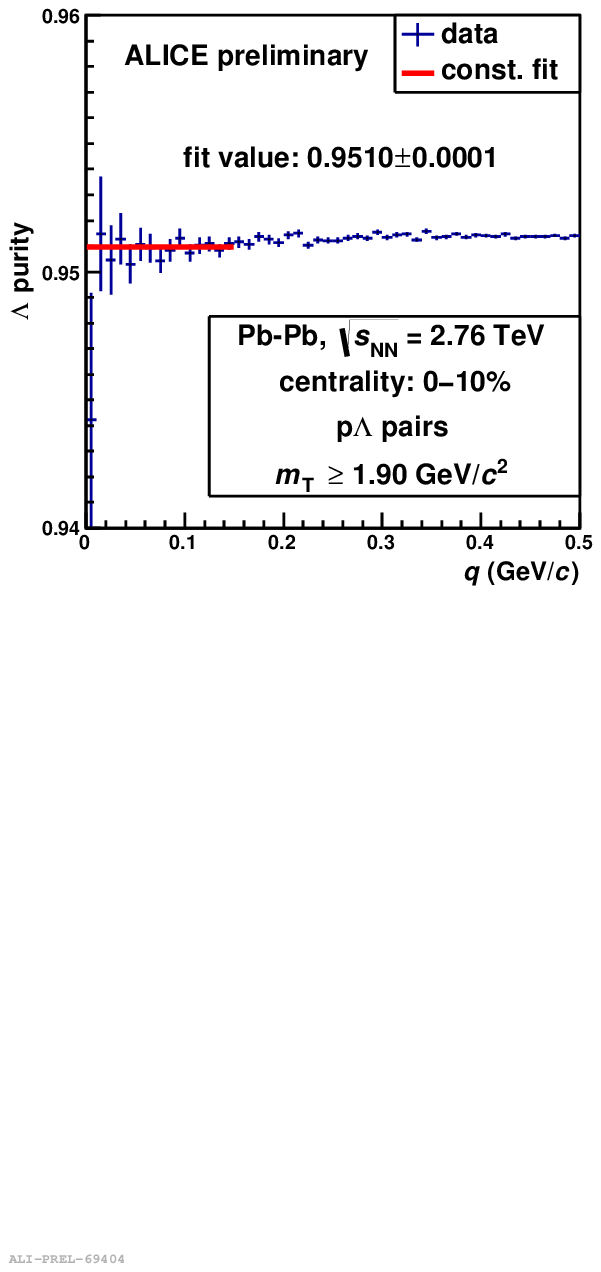}
    \end{array}$
  \end{center}
  \vspace{-0.5cm}
  \caption{Left: Distribution of the transverse distance of closest of approach
    of protons to the primary vertex in data with a fit utilizing
    Monte-Carlo templates in an exemplary phase-space bin.
  Right: \Lam\ pair purity for pairs with $\mT \ge 1.9~\gevc$
  for the 10\% most central events with a constant fit in the region
  $0.0 \le q~(\gevc) < 0.15$.}
  \label{fig:proDca-lamPairPur}
\end{figure}

The p\Lam\ strong interaction is limited in its range to a few
fm. Hence, any product from electro-magnetic or weak decays will not
contribute to the excess seen at small relative momenta. The
contamination in the proton sample mostly stems from the decay
$\Lam \rightarrow \mathrm{p}_{\mathrm{dec}}\uppi^-$. 
Since no significant
\Lam\Lam\ correlation was seen by the STAR
Collaboration~\cite{StarLamLam-HYP2012}, the decay proton will also
not carry any residual correlation~\cite{ResCorrI, ResCorrII,
  ResCorrIII} from the mother 
particle.\footnote{Recent data from ALICE~\cite{JaiWPCF2014} and
  STAR~\cite{STAR-LamLamPRL} suggest a slight \Lam\Lam\ 
  anti-correlation. The momentum released in the \Lam\ decay
  will wash this correlation out; the small fraction of
  $\mathrm{p}_{\mathrm{dec}}\Lam$ pairs in the p\Lam\
  sample will make it a tiny, likely negligible, correction.} The
determination of the amount of this uncorrelated feed-down allows for
a correction of the p\Lam\ correlation function with the
feed-down fraction 
$f_{\mathrm{p}} = \frac{\operatorname{non-primary}}{\operatorname{all}}$. 
The outstanding 
performance of ALICE allows to determine this feed-down fraction in
the proton sample directly from the data via the distance of closest
approach (DCA) of the track extrapolation to the primary
vertex. The two-dimensional (transverse and longitudinal) DCA
distribution was obtained differentially in rapidity, transverse
momentum, and centrality 
for the data, as well as for templates from Monte-Carlo
simulations for primary protons\footnote{According to the common ALICE
  definition, primary protons include decay products, except
  products from weak decays of strange hadrons.}, protons from weak 
decays, and protons from an interaction with the detector material.
The distinct shapes of the templates -- almost flat in DCA for
the material contribution, wide for the weak decays, and peaked for
the primaries -- allow to disentangle the different origins of the
charged nucleons by fitting the templates to the data, as pictured in
Fig.~\ref{fig:proDca-lamPairPur} (left). Selecting only 
tracks with a DCA smaller than 1~mm in the transverse and smaller than
1.5~mm in the longitudinal direction enhances the primaries in the sample.
In the exemplary phase-space bin $0.0 \le y < 0.25 $ and $1.0 \le
\pT~(\gevc) < 1.5$, the feed-down fraction $f_{\mathrm{p}}$ totals to 15\%. This
arises from the amount of particles which are from weak
decays and from material of 15\% and less than 1\%,
respectively. 

Also the \Lam\ sample is contaminated by feed-down. The contributing,
dominating weak decays are 
$\Xii^0\rightarrow \Lam \uppi^0$ (BR 99.5\%), 
$\Xii^-\rightarrow \Lam \uppi^-$ (BR 99.9\%),
$\Ome^-\rightarrow \Lam \mathrm{K}^-$ (BR 67.8\%), and
$\Ome^-\rightarrow \Xii (\rightarrow \Lam \uppi) \uppi$ (BR
32.2\%).
The reconstruction efficiencies with the chosen \Lam\ selection criteria
in this analysis for all $\mathrm{Y} = \Xii^0, \overline{\Xii}{}^0, \Xii^-,
\overline{\Xii}{}^+, \Ome^-$, and $\overline{\Ome}{}^+$ were
obtained three-dimensionally in $\pT^{\mathrm{Y}} \rightarrow (\pT^\Lam,
y^\Lam)$ for each centrality class of~\cite{ALICE-Multistrange}
from a Monte-Carlo simulation.  Using the \pT-differential spectra of
$\Xii^-, \overline{\Xii}{}^+, \Ome^-$, and $\overline{\Ome}{}^+$
measured by ALICE~\cite{ALICE-Multistrange} for several centrality
classes and assuming isospin symmetry for the unmeasured $\Xii^0$ and 
$\overline{\Xii}{}^0$, enabled us to determine the fraction of \Lam\
hyperons from weak decays.

The $c\tau = 22$~pm of the electro-magnetic decay
$\Sig^0\rightarrow \Lam \upgamma$ (BR 100\%) is much larger
than the range of the 
strong interaction, but also too small to be resolved experimentally.
In~\cite{pSigma0} it was found that the final-state interaction in
the p$\Upsigma^0$ channel is much smaller than in the p\Lam\ 
system. 
Thus, a significant fraction of \Lam\ come from $\Upsigma^0$ and are
uncorrelated with primary protons. We use 
the fraction of \Lam\ from $\Upsigma^0$ -- properly taking into
account all strongly decaying resonances --
from the thermal model of~\cite{Therminator}, while systematically
considering the study in~\cite{Andronic} with an
additional variation in the freeze-out temperature, and the results
of~\cite{Becattini} with the value at the kinetic and chemical
freeze-out. All models give that about 30\% of the \Lam\ originate
from a electromagnetic decay. We unite the fraction from weak and
electromagnetic hyperon decays in the feed-down fraction
$f_\Lam$. 

\section{Corrections and Results}
The raw p\Lam\ correlation as a function of the momentum
difference is obtained as the ratio of pairs from real events over
those reconstructed from mixed events for
three centrality divisions and up to four \mT\ classes for each
centrality class. Dealing with
non-identical particles, we use the generalized momentum difference
introduced in~\cite{LednickyGenMom} by R.~Lednick\'y 
$\tilde{q} = |q - P(qP)/P^2|, q=p_1-p_2, P = p_1+ p_2$,
where $p_1$ and $p_2$ are the momenta of the particles. In the
following, we omit the tilde. The overall pair purity factorizes:
$\mathrm{pur}(q,\mT) =
\mathrm{pur}_\Lam (q,\mT) \cdot
\left( 1-  f_{\mathrm{p}}(q,\mT) \right) \cdot
\left( 1-  f_\Lam (q,\mT) \right)$. The \Lam\ pair purity
$\mathrm{pur}_\Lam  (q,\mT)$ is shown exemplary in
Fig.~\ref{fig:proDca-lamPairPur} (right) for the 10\% most central
events and the highest \mT\ class, $\mT \ge 1.9~\gevcc$. Its value of 95.1\% is
constant over the region of interest $q < 0.15~(\gevc)$ and
beyond. It is obtained by looking up the
single-particle \Lam\ purity in the $(y,\pT)$-differential histograms
for the given centrality; the same holds for $f_{\mathrm{p}}$ and $f_\Lam$
accordingly. All
impurities constitute -- as discussed -- an uncorrelated
background. Hence, the corrected correlation function can be attained
by scaling the raw correlation function with the inverse pair purity:
\begin{equation}
  C_2^{\mathrm{corr.}} (q,\mT) = 
  \left( \frac{1}{\mathrm{pur}(q,\mT)} \cdot
    (C_2^{\mathrm{raw}} (q,\mT) -1)\right) +1.
  \label{eq:C2corr}
\end{equation}

\begin{wrapfigure}{R}{0.62\textwidth}
  \vspace{-0.5cm}
  \hspace{-0.02\textwidth}
  \includegraphics[width=0.7\textwidth]{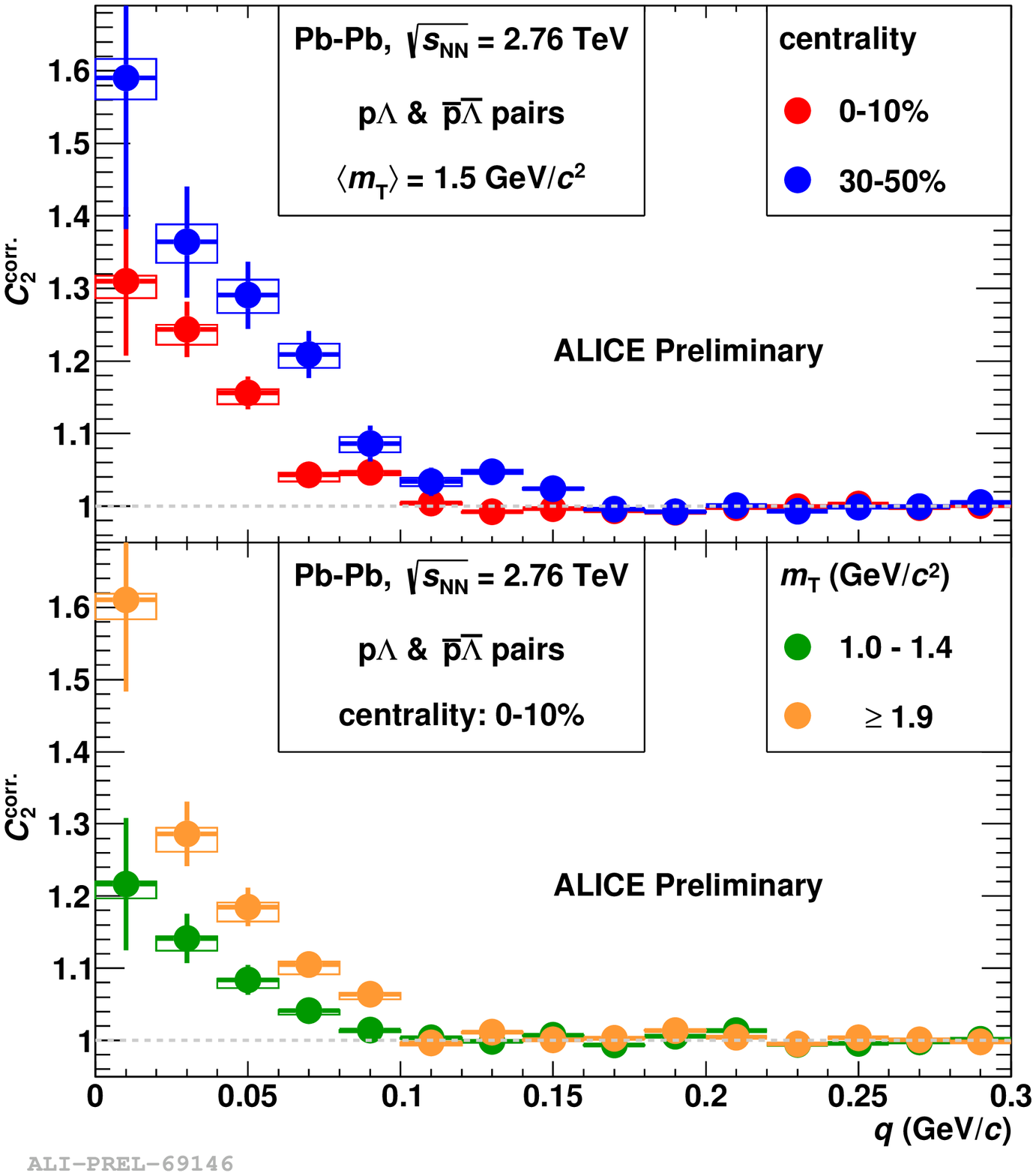}
  \caption{Exemplary p\Lam\ and
    $\overline{\mathrm{p}}\overline{\Lam}$ correlation function
  corrected for weak and electromagnetic decays. Centrality dependence
  at the top, \mT\ dependence at the bottom.}
  \label{fig:C2Corr}
\end{wrapfigure}
The effect of the finite momentum resolution was studied with a
Monte-Carlo simulation. 
For small relative momenta,
\ie\ $(q_{\mathrm{rec}}+q_{\mathrm{gen}})/2 \le 0.1~\gevc$,
 where $q_{\mathrm{rec}}$ is the reconstructed and $q_{\mathrm{gen}}$ is the
generated relative momentum, the deviation, quantified as
$(q_{\mathrm{rec}} - q_{\mathrm{gen}})/\sqrt{2}$, can be parametrized with a 
Gaussian with a mean of $-0.32\pm0.02$ and
$-0.38\pm0.08$~\mevc\ and a width of $7.26\pm0.02$ and
$7.13\pm0.07$~\mevc\ for the 0--10 and 30--50\% most central
events. In an \mT-differential study, the width turned out to slightly
increase by less than 2~\mevc\ for higher \mT.

Fig.~\ref{fig:C2Corr} shows a set of exemplary correlation functions
corrected via Eq.~\ref{eq:C2corr}, \ie\ remedied for the uncorrelated background
coming from weak and electro-magnetic decays.
The correlation functions for pairs of particles and pairs of 
anti-particles were merged
following the recipe of the Particle Data Group~\cite{PDG-merge}.
The systematic errors
include a variation of the correction for the electro-magnetic
feed-down, a change in the hyperon input spectra for the correction of
the \Lam\ from weak decays, an altered momentum resolution,
normalization, invariant mass selection of the \Lam, changed
DCA cuts on the proton, a varied two-track resolution cut, and
dominantly the uncertainty on the p\Lam\ interaction when
fitting the data.
The top panel shows the centrality dependence, \ie\ the 0--10\% most
central events in red and the 30--50\% most central events in
blue; both samples have a mean transverse mass of 1.5~\gevcc. One
clearly sees the expected effect of an increased width and 
height of the excess at small relative momenta, which translates into a
smaller source size, for the more peripheral collisions. The bottom panel
shows the dependence on \mT\ with the 
green points representing $1.0 \le \mT~(\gevcc) < 1.4$ resulting in $\langle \mT
\rangle$ = 1.27~\gevcc\ and the orange
symbols depicting $\mT \ge 1.9~\gevcc$ yielding $\langle \mT \rangle$ =
2.18~\gevcc\ for the 0--10\% most central events. Also here, a clear 
ordering is apparent with the higher \mT\ giving evidence for a
smaller source than at lower \mT, matching the expectation within the
hydrodynamic picture outlined in the introduction. Note that the
correlation function with $\langle \mT \rangle = 2.18~\gevcc$ exceeds
the $\langle \mT \rangle$ of any previous measurement from~\cite{NA49plam} at
the SPS, \cite{Starpp} at RHIC or \cite{ALICE-Femto-QM12,
  ALICE-Femto-HQ12} at the LHC. 
\section{Summary}
The supreme performance of \ALICE\ makes it possible to collect very pure
samples of protons and \Lam\ with rich statistics. We obtained p\Lam\ 
correlation functions multi-differentially 
in centrality and transverse mass. They were corrected for
misidentification and contamination from weak and electromagnetic
decays, employing data-driven methods to quantify the
impurities. The correlation functions presented here represent the
largest \mT\ reach of any femtoscopic measurement, with result being
shown for $ \mT \ge 1.9~\gevcc$. The conveyed centrality dependence
of the p\Lam\ correlations exhibits the expected behavior of a
smaller source for more peripheral collisions. 
The communicated
\mT-differential correlations, spanning a range in $\langle \mT
\rangle$ of more than 0.9~\gevcc, display the decrease in source size
with \mT, qualitatively agreeing with hydrodynamic predictions, out to highest 
$\langle \mT \rangle = 2.18~\gevcc$.

\end{document}